# The Factor 2 in Fundamental Physics

## Peter Rowlands

*IQ Group and Science Communication Unit, Department of Physics, University of Liverpool, Oliver Lodge Laboratory, Oxford Street, Liverpool, L69 7ZE, UK. e-mail prowl@hep.ph.liv.ac.uk and prowl@csc.liv.uk*

*Abstract.* A brief history is given of the factor 2, starting in the most elementary considerations of geometry and the kinematics of uniform acceleration, and moving to relativity, quantum mechanics and particle physics. The basic argument is that in all the significant cases in which the factor 2 or ½ occurs in fundamental physics, whether classical, quantum or relativistic, the same physical operation is taking place.

## 1 Geometry and kinematics

We probably first come across the factor 2 in the formula for the triangle:

$$\text{area} = \frac{\text{length of base} \times \text{perpendicular height}}{2}.$$

This is an ancient formula, well-known to Egyptian, Babylonian and Chinese mathematicians. In the case of a right-angled triangle, it is clearly created by bisecting a rectangle along a diagonal. If we now take this as representing a straight-line graph, of, say, velocity against time, under a uniform acceleration, the area under the graph becomes the distance travelled. For an object increasing its velocity uniformly from 0 to a value $v$ in time interval $t$, the area under the graph, or distance travelled, using the triangle formula, becomes $vt/2$. By comparison, if the object had travelled at steady speed $v$ throughout the time interval $t$, the distance travelled would be the area of the rectangle under the horizontal straight line representing steady $v$, that is, $vt$. In effect, the factor 2 distinguishes here between steady conditions and steadily *changing* conditions.

It was by this means that the factor first entered into physics from pure mathematics, via the Merton mean speed theorem, evolved in fourteenth-century Oxford. This result, which ultimately proved to be the foundation theorem of modern dynamics, showed that the total distance moved by a body during uniform acceleration was the same as that covered during the same time interval by a body travelling uniformly at the speed measured at the middle instant of the accelerated motion. In more modern terms, the total distance travelled under uniform acceleration must equal the product of the mean speed and the time. Mathematically, if we start with initial speed $u$ and steadily accelerate to a final speed $v$ over the time interval $t$, then the total distance travelled will be given by

$$s = \frac{(u+v)t}{2}.$$

This is, of course, identical to the value we would obtain from our straight line graph if we took the area under the graph between $u$ and $v$ as the sum of a rectangle ($ut$) and a triangle (($v-u$)$t/2$), and reduces to $vt/2$ when $u=0$.

If we additionally use the definition of uniform acceleration, $a=(v-u)/t$, we obtain the well-known equation for uniformly accelerated motion:

$$v^2 = u^2 + 2as,$$

which reduces to $v^2 = 2as$, when $u=0$. If we now apply this to a body of mass $m$, acted on by a uniform force $F=ma$, we find the work done over distance $s$ is equal to the kinetic energy gained

$$Fs = mas = \frac{mv^2}{2} - \frac{mu^2}{2},$$

which reduces to $mv^2/2$ if we start at zero speed. Using $p = mv$ to represent momentum, it is convenient also to express this formula in the form $p^2/2m$. Of course, this formula applies more generally than in the case of purely uniformly accelerated motion, and we may derive the more general formula for nonuniformly accelerated motion by a simple integration of force ($dp/dt$) over displacement:

$$\int \frac{dp}{dt} ds = \int mv\,dv = \frac{mv^2}{2}.$$

However, the example of uniform acceleration, treated graphically, shows, in a strikingly simple manner, the origin of the factor of 2 in a process of averaging over changing conditions. In this context, the factor 2 relates together the two main areas of dynamical physics: those of accelerated and unaccelerated straight line motion. For the case of zero initial velocity, the distance travelled under uniform acceleration can be represented as the area of a triangle on a $v$-$t$ graph compared with the rectangle representing uniform velocity. In effect, a steady increase of velocity from 0 to $v$ requires an averaging out which halves the values obtained under steady-state conditions.

**2 Kinetic and potential energy**

It is in precisely the same way that the factor 2 makes its appearance in molecular thermodynamics, quantum theory and relativity. It is, in a sense, the factor which relates the continuous aspect of physics to the discrete, and, as both these aspects are required in the description of any physical system, the factor acquires a universal relevance. The most obvious classical manifestation is the fact that *two* types of energy equation are commonly used in physics, both of which are expressions of the general law of conservation of energy, but each of which expresses this fundamental truth in a subtly different way. One is the potential energy equation representing steady-state conditions, which applies wherever there is no overall change in the energy distribution; the kinetic energy equation, on the other hand, requires a redistribution of energy within a system while maintaining the overall principle of energy conservation.

In a typical example, we apply the potential energy equation to the case of a planet in a regular gravitational orbit. So, the force equation

$$\frac{mv^2}{r} = -\frac{GMm}{r^2}$$

leads to a potential energy relation

$$mv^2 = -\frac{GMm}{r}.$$

On the other hand, the changing conditions involved in the escape of a body of mass *m* from a gravitational field require a kinetic energy equation of the form

$$\frac{mv^2}{2} = \frac{GMm}{r}.$$

Significantly, Newton, despite having no word or expression equivalent to the modern term 'energy' or to any particular form of it, used both these equations in his *Principia*, in the more general forms applicable to any force.[1] Book I, Proposition XLI, a version of what came to be known as the '*vis viva*' integral, is applied to finding the paths taken by bodies subject to any type of centripetal force; this is a classic case of the potential energy equation. Proposition XXXIX, on the other hand, which considers the velocity of a rising or falling body produced by the action of an arbitrary force, is a kinetic energy equation, showing that the work done, or the integral of force over distance, in unresisted motion, is equal to the change in kinetic energy produced ($\Delta W = \Delta(mv^2/2)$).

Numerically, we observe that the potential energy term is twice the value of the kinetic. We recognise here, of course, that this is a special case of the virial theorem relating the time-averaged potential and kinetic energies, $\bar{V}$ and $\bar{T}$, in a conservative system governed by force terms inversely proportional to power *n* of the distance, or potential energy terms inversely proportional to power *n*–1. That is:

$$\bar{V} = \frac{(1-n)}{2}\bar{T}.$$

The virial theorem, in effect, gives us a relationship between the energy term relevant to constant conditions (potential energy) and that obtained under conditions of change (kinetic energy). For two special cases – constant force and inverse-square-law force – $\bar{V}$ will be numerically equal to $2\bar{T}$, for in these cases $n$ is, respectively, equal to 0 and 2. Such forces, in fact, are overwhelmingly predominant in nature, as they are a natural consequence of three-dimensional space. In many cases, then, the factor 2 becomes the direct expression of the relationship between potential and kinetic energies.

## 3 Kinetic theory of gases

The fact that two apparently contradictory equations can both be said to illustrate the general principle of the conservation of energy can be easily explained if we consider the kinetic energy relation to be concerned with the action side of Newton's third law, while the potential energy relation concerns both action and reaction. Because of the necessary relation between them, each of these approaches is a proper and complete expression of the conservation of energy. However, circumstances generally dictate which of the two is the most appropriate to use. A good illustration of the connection is given by an old proof of Newton's of the $mv^2/r$ law for centripetal force, and hence of the formula $mv^2$ for orbital potential energy.[2] This had the satellite object being 'reflected' off the circle of the orbit, first in a square formation, and then in a polygon with an increasing number of sides, becoming, in the limiting case, a circle. Here, the momentum-doubling action and reaction by the imagined physical reflection produces the potential energy formula, as well as demonstrating the relation between the conservation laws of linear and angular momentum.

Another significant case is the derivation of Boyle's law, or the proportionality of pressure ($P$) and density ($\rho$) in an ideal gas, from what is often described as the 'kinetic' theory. Contrary to what is often stated in elementary textbooks, the kinetic behaviour of gas molecules is in no direct way responsible for Boyle's law. The derivation involves a doubling of momentum as the ideal gas molecules reflect off the walls of the container, of the same kind as Newton assumed in his imaginary reflections under centripetal acceleration. The factor 2 thus introduced is then immediately removed by the fact that we have to calculate the average time between collisions ($t = 2a/v$) as the time taken to travel *twice* the length of the container ($a$). The average force then becomes the momentum change / time = $2mv/t = mv^2/a$, and the pressure due to one molecule in a cubical container of side $a$ becomes $mv^2/a^3$, or $mv^2/V$ (volume), leading for $n$ molecules to the pressure-density relationship.

The incorporation of momentum-doubling means that both action and reaction are included in the system under consideration, thereby creating a steady-state dynamics with positions of molecules constant on a time-average. Taking into account the three dimensions between which the velocity is distributed, the ratio of pressure and density ($P/\rho$) is derived from the *potential energy* term $mv^2$ for each molecule and is equal to one third of the average of the squared velocity, or $\overline{c}^2/3$.

Strictly speaking, this result has nothing whatsoever to do with any dynamical model of gas molecule behaviour. Newton derived exactly the same result, assuming, *for purely mathematical purposes*,[3] that gas molecules could be considered as stationary objects exerting a repulsive outward force on each other in inverse proportional proportion to their distance apart, and a gas in steady state exerts a pressure in all directions which is exactly the same as the molecules being considered stationary on a time average and exerting a force inversely proportional to their mean distance apart, or, for a fixed mass of gas, to the length of the container. (It is not, of course, necessary to assume that this is due to a physical interaction between the molecules.)

We only bring in kinetic behaviour when we relate the average kinetic energy of the molecules to the temperature of the gas; but there is no 'derivation' involved because temperature is not defined independently of the kinetic energy, and we make this definition by an *explicit* use of the virial theorem, to find the *unknown* average kinetic energy from the *known* potential energy. We find that the potential energy of each individual molecule is $kT$ for each degree of freedom, and, in total, 3$kT$. However, by applying the virial theorem to the results of the potential energy calculations, we can relate the dynamical behaviour of an ideal gas to the kinetic energy expression (3$kT/2$) for its individual molecules. In effect, the derivation of Boyle's law assuming dynamical gas molecules was merely an operational convenience; for pressure terms of any kind, whatever their origin, are an expression of the action of force or potential energy. That is, it is a purely formal matter whether we describe the gas in terms of the average kinetic energy of the individual molecules or an equivalent averaged-out potential energy of the gas as a whole. A gas in steady state is equivalent to a system with constant expansive force in all directions, and a system of this kind necessarily requires a virial relation of the form

$$\overline{V} = 2\overline{T}.$$

between the time-averaged potential and kinetic energies.

It is interesting, incidentally, that Newton's earliest derivation of the centripetal force law ($mv^2/r$) (prior to the geometrical proof discussed above)[2] involved essentially the same collision process (involving aether particles) as is now used in the kinetic theory of gases, with force calculated as the product of the change in momentum in a particle due to impact and the rate at which collisions take place, the collision rate being found by dividing the particle velocity by the distance travelled between collisions.

## 4 Radiation pressure

Consideration of material gases leads us on to the subject of photon 'gases', as first considered by Einstein in deriving radiation pressure, following the earlier, classical, calculation by Boltzmann. Remarkably, the expressions for photon gases are *identical in form* to those for material gases, even though the photon gas is a relativistic system, unlike the material gas. In exact parallel to the expression for a material gas, the radiation pressure of a photon gas within a fixed enclosure is found to be one third of the energy density of radiation, that is:

$$P = \frac{1}{3}\rho c^2.$$

In this context, the photon behaves in exactly the same way as a material particle, and, because the system is in steady state, the energy term $mc^2$ behaves as *potential*, not kinetic, energy, exactly as its form would suggest. The photons are reflected off the walls of the container in the same way as the material gas molecules, although this time we can also consider the process as involving absorption and re-emission, There is thus no mysterious 'relativistic' factor at work here – as suggested by some authors who see $mc^2$ for the photon as a 'kinetic' energy replacing the term $mv^2/2$ used for material particles; $mc^2$ is simply a reflection of the potential nature of the photon's total energy.

The whole point of Einstein's introduction of the formula $E = mc^2$ to represent the photon's total energy (and, by analogy, that of true material particles) was to preserve the *classical* laws of conservation of mass and conservation of energy. As Einstein himself was well aware, the total energy equation $E = mc^2$ cannot be derived, by deductive means, from the postulates of relativity; all that can be demonstrated is the *change of energy* formula $\Delta E = \Delta mc^2$. It is merely an act of faith to extend this formula to the more general expression. This is because the total energy term occurs only as a constant of arbitrary value in the integration of the relativistic expression for rate of energy change:

$$\frac{dT}{dt} = \mathbf{F.v}.$$

In addition, the presence of $mc^2$ in the relativistic kinetic energy equation, which emerges as the solution to this integral:

$$T = \frac{mc^2}{(1 - v^2/c^2)^{1/2}}$$
$$= mc^2 + \frac{mv^2}{2} + ...$$

contradicts the well-established principle that special relativistic equations lead to classical ones when $v \ll c$. In principle, we could add any constant of integration to the equation. Adding $mc^2$, for example, would remove the anomalous term entirely and make the expression, for $v \ll c$, identical with the classical one, as would normally be required. It has been found convenient, in relativity, however, to take the constant of integration as 0, because this allows a convenient definition of a 4-vector momentum, and then to find a 'physical' meaning for the added term $mc^2$.

Writers who have investigated Einstein's own arguments and who demonstrate the validity of his derivation of the equation $\Delta E = \Delta mc^2$ point to the arbitrary, though physically reasonable, nature of its extension to a body's total mass. Stachel and Toretti, for example, state that: 'The final conclusion that the entire mass of a body is in effect a measure of its energy, is of course entirely unwarranted by Einstein's premises';[4] and they quote Einstein as follows: 'A mass $m$ is equivalent – insofar as its inertia is concerned – to an energy content of magnitude $mc^2$. Since we can arbitrarily fix the zero of (the total energy), we are not even able to distinguish, without arbitrariness, between a 'true' and an 'apparent' mass of the system. It appears much more natural to regard all material mass as a store of energy.'[5]

Of course, what is arbitrary in special relativity need not be arbitrary in other contexts; if an idea is 'physically reasonable' or 'natural', it must be explicable in terms of some definite physical principles; and if mass is to be considered as a 'store' of energy, then this principle must be related to the idea of mass as a specifically potential form of energy. No problem, therefore, arises if we recognise that $mc^2$ has a classical, *as well as relativistic*, meaning. Its structure is clearly that of a classical potential energy, which is precisely what we would expect *total* energy to be, and it was introduced to preserve a classical conservation law. Like many other things in relativity (the Schwarzschild radius, the equations for the expanding universe, the gravitational redshift, the spin of the electron), the expression does not arise from the theory of relativity itself but is a more fundamental truth which that theory has uncovered.

**5 The classical potential energy of the photon**

The number 2 has frequently been described as a 'relativistic' factor separating relativistic and nonrelativistic cases, but it is no such thing. It would be extraordinary if relativistic conditions should somehow conspire exactly to halve or double significant classical quantities. Relativistic factors are typically of the form $\gamma = (1 - v^2/c^2)^{-1/2}$, suggesting some gradual change when $v \to c$. It makes no physical sense to suppose that the transition involves discrete integers. Certainly, $\Delta E = \Delta mc^2$ is a relativistic equation because it incorporates the $\gamma$ factor in the $\Delta m$ term, but $E = mc^2$ is not, even though it took relativity to discover its application to material particles. $mc^2$ is a potential energy term in classical physics which has the same effect as the equation $E = mc^2$ in relativistic physics, and the effects which depend only on $E = mc^2$ and not specifically on the 4-vector combination of space and time can be derived by classical approaches entirely independent of any concept of relativity. In fact, in the special case of light photons – or light 'corpuscles' in the older terminology – potential energy terms of the form $mc^2$, or their equivalent, have been regularly used since the seventeenth century in a variety of classical contexts, and are still so used in specialised calculations.

Newton, for example, in examining atmospheric refraction in 1694, conceived the bending of light as equivalent (in our terms) to a change in potential energy from $mc^2$ as a result of a constant refracting field analagous to the gravitational field **g** at the Earth's surface.[6-8] The equation he used was effectively the same as our steady-state potential energy (or '*vis viva*') equation for a circular gravitational orbit, slightly modified for the elliptical case. Ordinary refraction he treated as a process analagous to gravitational orbital motion, subject to a force $mc^2/r$, and, by implication, a potential energy $mc^2$, analogous to the gravitational orbital force $mv^2/r$ and gravitational potential energy $mv^2$. This analogy is possible because the constancy of the velocity of light ensures that the optical system is 'steady-state' and that its potential energy term is numerically equivalent to that in the inverse-square-law gravitational system.

As a result of this second calculation, Newton was able to write in a draft of Query 22/30 for the *Opticks* that: '…upon a fair computation it will (be) found that the gravity of our earth towards the Sun in proportion to the quantity of its matter is above ten hundred million of millions of millions of millions of times less then the force by wch a ray of light in entering into glass or crystal is drawn or impelled towards the refracting body.…For the velocity of light is to the velocity of Earth in Orbis magnus as 58 days of time (in which) the Earth describes the (same space –); that is an arch equal to the radius of its orb to about 7 minutes, the time in wch light comes from (the Sun) to us; that is as about 12,000 to 1. And the radius of the curvity of a ray of light during it(s) refraction at the surface of glass on wch it falls very obliquely, is to the curvity of the earth Orb, as the radius of that Orb to the radius of curvature of the ray or as above 1,000,000,000,000,000,000 to 1. And the force wch bends the ray is to the force wch keeps the earth or any Projectile in its orb or line of Projection in a ratio compounded of the duplicate ratio of the velocities & the ratio of the curvities of the lines of projection.'[9-10]

In another calculation in the same manuscript, Newton took the radius of the Earth's orbit as 69 million miles (based on a solar parallax of 12 seconds) and the radius of curvature of the path of a light particle as $10^{-6}$ inches. Assuming that the light from the Sun takes 7.5 minutes to reach the Earth and that in this time the Earth would have travelled 6197 miles, he found the ratio of the forces to be about $5 \times 10^{26}$. The centripetal force calculation used in Newton's studies of refraction is an illustration of the power of the virial theorem, and must give the correct numerical energy relation whether or not the description of the force is 'correct'. The constraints which need to be applied to find the true nature of the vector force are not required to find the numerical value of the scalar energy.

A quite different use is found in Newton's formula for the velocity of waves in a medium, in terms of elasticity or pressure ($E$) and density ($\rho$), which he applied to both light and sound. [11] Essentially, Newton's formula

$$c = (E/\rho)^{1/2}$$

is an expression of the fact that the potential energy of the system of photons, or gas molecules in the case of sound ($mc^2$), is equal to the work done at constant pressure as a product of pressure and volume. The application to light, or at least to its medium of propagation, occurs in the calculation in the published Query 21 of the ratio between the elasticities per unit density of a proposed electro-optic 'aether' and atmospheric air, and the manuscript evidence shows that this was linked also with the calculation of the force of refraction, occurring immediately after the final version of that calculation in the manuscript. [10]

Newton's elasticity of the aether is what we would call energy density of radiation ($\rho c^2$), which is related by Maxwell's classical formula of 1873 to the radiation pressure, and the ratio he calculates is, in effect, the ratio of the energy per unit mass of a particle of light to the energy per unit mass of an air molecule, as manifested in the transmission of sound. Now, the molecular potential energy per unit mass for air may be calculated from the kinetic theory of gases ($PV = RT$) at about $1.8 \times 10^5$ J kg$^{-1}$ when $T = 300$K. Since the energy per unit mass for a light photon is $9 \times 10^{16}$ J kg$^{-1}$, the ratio is $5 \times 10^{11}$, which is comparable with Newton's $4.9 \times 10^{11}$ minimum in Query 21. Light, of course, will always give 'correct' results for such a calculation when travelling through a vacuum, because in such circumstances, there is no source of dissipation, and the virial relation takes on its ideal form. Newton's formula for calculating the velocities of waves in a medium is thus another perfect illustration of an application of the virial theorem, and it is because it is such a perfect illustration that it works in a case where the model of interaction with matter no longer applies. This is why the elasticity of light is precisely the same thing as its energy density. Although this does not apply as exactly to sound, where (as Laplace later showed) the 'elasticity' constant needs to be calculated from the adiabatic, rather than the isothermal, value, the correction factor is relatively small in order of magnitude terms (20%).

## 6 The gravitational bending of light

Interestingly, though light in free space has velocity $c$, and, therefore, no rest mass or kinetic energy, *as soon as you apply a gravitational field*, the light 'slows down', and, at least *behaves* as though it can be treated as a particle with kinetic energy *in the field*. This is precisely what happens when we use the standard Newtonian escape velocity (or kinetic energy) equation

$$\frac{mv^2}{2} = \frac{GMm}{r}$$

to derive the Schwarzschild limit for a black hole, by purely classical means, as was done in the eighteenth century by Michell and Laplace.[12-13] Assuming $v \to c$, we derive

$$r = \frac{2GM}{c^2}$$

and *there is no transition to a 'relativistic' value*.

A classic case of applying a kinetic energy-type equation to light, is the classical derivation of the double gravitational bending, an effect normally thought to be derivable only from the general relativistic field equations. The double bending of light in a gravitational field has been a *cause celêbre* since Eddington used it to establish Einstein's theory in the eclipse expedition of 1919.[14] We have since that time been repeatedly assured that the double bending is a relativistic effect, and that 'Newtonian' calculations, using the principle of equivalence, yield only half the correct value, although several authors have put forward demonstrations of the double deflection based only on special relativity.[15-18]

The 'Newtonian' calculation takes its origin from Soldner, who, we are told, in a paper of 1801,[19-21] investigated the gravitational deflection of light by a massive body, using the standard '*vis viva*' theorem or potential energy equation (modified for a hyperbolic orbit), according to the expression:

$$mc^2 = \frac{GMm(e-1)}{r},$$

with $e$ taken as the eccentricity of the hyperbolic orbit. Since $1 \ll e$, the half-angle deflection becomes

$$\frac{1}{e} = \frac{GM}{c^2 r},$$

and the full angle deflection (that is, in and out of the gravitational field)

$$\frac{2}{e} = \frac{2GM}{c^2 r}.$$

General relativity, however, finds

$$\frac{2}{e} = \frac{4GM}{c^2 r},$$

and it was the supposed experimental realisation of this result which allowed Eddington to claim that he had 'overthrown' Newtonian physics.

However, Soldner did not use the potential energy equation. He used the kinetic energy equation,

$$\frac{mc^2}{2} = \frac{GMm(e-1)}{r},$$

on the basis of Laplace's prior employment of it in calculating the black hole radius, and he would have obtained the 'correct' total deflection if he had used the double angle in calculating his integral! This was, indeed, correct procedure, for the deflection of a photon coming past the Sun's edge from infinity is a case of an orbit in the process of formation, and not an orbit in steady state. It is the reverse of the process of creating an orbit by escaping from the confining field, modified, of course, by the hyperbolic, rather than circular or elliptical, orbit produced by the immense relative speed of the light photon. The significant consideration is that, on its immensely long journey prior to its coming close to the gravitational field of the Sun, the photon's velocity was not determined by the Sun's gravitational field, and the direction of its deflection is perpendicular to this. The classical equation used by Eddington was the one specified for steady-state conditions, whereas light-bending is surely an example of energy exchange.

We should not be surprised that a purely classical calculation of the light-bending is possible in this way. In principle, relativity theory does not produce different energy equations to classical physics; it merely corrects our naïve understanding of what are steady-state and what are changing conditions. The photon, in particular, provides an instance in which we would *expect* relativistic equations to coincide with classical ones. Photon energy, after all, is field energy and has no material component; the photon mass is, therefore, defined in terms of a pre-existing classical energy equation and does not provide a source of independent information which can be used to distinguish between classical and relativistic conditions.

The use of a 'kinetic energy' expression $mc^2/2$ in the case of light bending does not, of course, imply that photon 'total energy' is of this form, or that there is any such thing as the 'kinetic energy' of a photon; $mc^2/2$ (as in the parallel case of the derivation of the Schwarzschild radius) is merely an expression of the action of the perturbing field. We never see this energy directly, for, whenever a photon interacts with matter (or is 'detected'), its 'independent' existence has ceased and the energy absorbed is purely the potential or total energy value $mc^2$. It is this aspect of the photon's existence that has led to the idea that the absence of the factor 2 is somehow a mysterious property of relativity not paralleled in classical physics.

The idea that a 'relativistic' correction (either special or general) 'causes' the doubling of gravitational effect is an illustration, not of the fact that the calculation has to be done in a relativistic way, but that relativity provides one way of incorporating the effect of changing conditions if we begin with the potential, rather than the kinetic, energy equation. Here, the potential energy equation typically produces the effect of gravitational redshift, or time dilation, while relativity adds the corresponding length contraction. So some authors have argued for the redshift being 'Newtonian' while the length-contraction or 'space-warping' is relativistic, while others claim that the reverse is true. It has also been claimed variously that the 'Newtonian' effect has to be added to that produced by the Einstein calculation of 1911, based on the equivalence principle (which also obtained only half of the correct value), or that the two effects are the same, and have to be supplemented by a 'true' relativistic effect, like the Thomas precession.[16,22-25] It is, of course, purely (classical) energy considerations which decide the issue. If the potential energy equation is used where the kinetic energy equation is appropriate, then (correct) physical reasons can be found for almost *any* additional term which doubles the effect predicted.

The true nature of the contributions made by different causes to the three relativistic predictions of redshift, light bending and perihelion precession has been obscured by the all-embracing nature of the general relativistic formalism, and it is too easily assumed that the effects can be derived only from the full field equations of general relativity. Comparison with classical predictions demonstrate that redshift and the time dilation components of light bending and perihelion precession depend only on the relation $E = mc^2$ and not on the 4-vector combination of space and time. The spatial components of the light bending and perihelion precession should then follow automatically from the application of 4-vector space-time without any need to apply the equivalence principle, any time dilation necessarily requiring an equivalent length contraction. However, since $mc^2$ in a field has a 'kinetic energy' equivalent, even special relativity is only an alternative approach to a calculation that must also be valid classically.[14,26-27]

In a historical context, although we have no direct calculation of the 'Newtonian' deflection of light from Newton himself, there is a related calculation of atmospheric refraction using the potential energy equation, similar to the one already mentioned.[6-8] Newton assumes a constant refracting field $f$ at a height $h$ above the Earth's surface, entirely analagous to the gravitational field $g (= GM / r^2)$. He then uses Proposition XLI, to calculate the resulting deflection into parabolic orbits of light rays entering the Earth's atmosphere. The assumption of parabolic orbits requires $mc^2$ to be equated to the potential energy term $mfr(1 + \cos \varphi)$, equivalent to the gravitational $GMm(1 + \cos \varphi)/r$, while the use of Proposition XLI is equivalent to a modification of $c^2$ by the factor $(1 - 2fh / c^2)$ in the same way as the principle of equivalence is used to modify $c^2$ by $(1 - 2gr / c^2)$ or $\gamma^{-2}$ in gravitational bending. Significantly, atmospheric refraction is still calculated in modern astronomical textbooks using the *old* corpuscular theory!

**7 The gyromagnetic ratio of the electron**

Relativity has also been assumed to be needed to explain the anomalous magnetic moment or, equivalently, the gyromagnetic ratio of a Bohr electron acquiring energy a magnetic field. According to 'classical' reasoning, it has been supposed, an electron changing its angular frequency from $\omega_0$ to $\omega$ acquires energy in a magnetic field **B** of the form

$$m(\omega^2 - \omega_0^2) = e\omega_0 rB,$$

leading, after factorization of ($\omega^2 - \omega_0^2$), to an angular frequency change

$$\Delta\omega = \frac{eB}{2mr}.$$

However, a relativistic effect (the Thomas precession, again) ensures that the classical $e\omega_0 rB$ is replaced by $2e\omega_0 rB$, leading to

$$\Delta\omega = \frac{eB}{mr}.$$

But relativistic and classical treatments coincide when, as with the light bending example, the *kinetic* energy equation is recognised as the one applied to changing conditions, *at the instant we 'switch on' the field*. Then, we automatically write

$$\frac{1}{2}m(\omega^2 - \omega_0^2) = e\omega_0 rB,$$

which is no more, in principle, than the equation of motion for uniform acceleration

$$v^2 - u^2 = 2as.$$

So, the Thomas precession is needed if we begin with the potential energy equation applicable to a steady state, but not if we apply the kinetic energy used for changing conditions.

**8 The Dirac equation**

The gyromagnetic ratio leads on naturally to the subject of electron spin. For this, we need to introduce the Dirac equation. Here it will be convenient to rewrite the Dirac equation,

$$(\gamma^\mu \partial_\mu + im)\psi = 0$$

or

$$(i\boldsymbol{\gamma}\cdot\mathbf{p} + m - \gamma_0 E)\psi = 0,$$

in a more algebraic form with the $\gamma$ matrices replaced by a combination of quaternion and multivariate 4-vector algebras.[28-36] Here, we write

$$\gamma_0 = i\boldsymbol{k} \; ; \; \gamma_1 = \mathbf{i}\boldsymbol{i} \; ; \; \gamma_2 = \mathbf{j}\boldsymbol{i} \; ; \; \gamma_3 = \mathbf{k}\boldsymbol{i} \; ; \; \gamma_5 = i\boldsymbol{j}.$$

The quaternion units, $i, j, k$, follow the usual multiplication rules for quaternions; the multivariate vector units, **i**, **j**, **k**, follow the multiplication rules for Pauli matrices:

| *vector units* | *quaternion units* |
|---|---|
| $\mathbf{i}^2 = 1$ | $i^2 = -1$ |
| $\mathbf{ij} = -\mathbf{ji} = i\mathbf{k}$ | $ij = -ji = k$ |
| $\mathbf{j}^2 = 1$ | $j^2 = -1$ |
| $\mathbf{jk} = -\mathbf{kj} = i\mathbf{i}$ | $jk = -kj = i$ |
| $\mathbf{k}^2 = 1$ | $k^2 = -1$ |
| $\mathbf{ki} = -\mathbf{ki} = i\mathbf{j}$ | $ki = -ik = j$ |

The reformulation is necessary to the understanding of how the Dirac equation relates to classical energy conservation rules, for, using it, we can easily derive the Dirac equation, via the Correspondence Principle. We take the classical relativistic energy-momentum conservation equation:

$$E^2 - p^2c^2 - m_0^2c^4 = 0,$$

and factorize using our quaternion-multivariate-4-vector operators to give:

$$(\pm kE \pm i\mathbf{i}\,\mathbf{p} + ij\,m_0)(\pm kE \pm i\mathbf{i}\,\mathbf{p} + ij\,m_0) = 0.$$

Adding an exponential term and replacing the left-hand bracket by quantum differential operators, we obtain

$$\left(\pm ik\frac{\partial}{\partial t} \pm i\,\nabla + ijm_0\right)\psi = 0,$$

where

$$\psi = (\pm kE \pm i\mathbf{i}\,\mathbf{p} + ij\,m_0)e^{-i(Et-\mathbf{p}\cdot\mathbf{r})}.$$

The four solutions possible with $\pm E, \pm \mathbf{p}$, may be represented by a column vector with the four terms:

$$(kE + i\mathbf{i}\,\mathbf{p} + ij\,m_0)$$
$$(kE - i\mathbf{i}\,\mathbf{p} + ij\,m_0)$$
$$(-kE + i\mathbf{i}\,\mathbf{p} + ij\,m_0)$$
$$(-kE - i\mathbf{i}\,\mathbf{p} + ij\,m_0),$$

representing a single quantum state. We can proceed to show that a spin 1 boson wavefunction (incorporating fermion-antifermion combination) is the sum of

$$( k E + i i\ \mathbf{p} + i j\ m_0)(-k E + i i\ \mathbf{p} + i j\ m_0)$$
$$( k E - i i\ \mathbf{p} + i j\ m_0)(-k E - i i\ \mathbf{p} + i j\ m_0)$$
$$(-k E + i i\ \mathbf{p} + i j\ m_0)( k E + i i\ \mathbf{p} + i j\ m_0)$$
$$(-k E - i i\ \mathbf{p} + i j\ m_0)( k E - i i\ \mathbf{p} + i j\ m_0)$$

while a spin 0 boson is the sum of

$$( k E + i i\ \mathbf{p} + i j\ m_0)(-k E - i i\ \mathbf{p} + i j\ m_0)$$
$$( k E - i i\ \mathbf{p} + i j\ m_0)(-k E + i i\ \mathbf{p} + i j\ m_0)$$
$$(-k E + i i\ \mathbf{p} + i j\ m_0)( k E - i i\ \mathbf{p} + i j\ m_0)$$
$$(-k E - i i\ \mathbf{p} + i j\ m_0)( k E + i i\ \mathbf{p} + i j\ m_0),$$

each multiplied by the usual exponential form in creating the wavefunction. The fermion wavefunction is effectively a nilpotent (a square root of 0), and the boson wavefunction a product of two nilpotents (each not nilpotent to the other). The multiplications here are scalar multiplications of a 4-component bra vector (composed of the left-hand brackets), representing the particle states, and a ket vector (composed of the right-hand brackets), representing the antiparticle states.

## 9 Electron spin from the Dirac equation

The conventional treatment of spin introduces the factor 2 through the property of noncommutation of vector operators. From the standard version of the Dirac equation, we obtain

$$[\hat{\sigma}, \mathcal{H}] = [\hat{\sigma}, i\gamma_0 \boldsymbol{\gamma} \cdot \mathbf{p} + \gamma_0 m].$$

where $\mathcal{H}$ is the Hamiltonian, or total energy operator, and

$$\hat{\sigma}_l = i\gamma_0 \gamma_5 \gamma_1, \text{ with } l = 1, 2, 3$$

and

$$i\gamma_0 \boldsymbol{\gamma} \cdot \mathbf{p} = i\gamma_0\gamma_1 p_1 + i\gamma_0\gamma_2 p_2 + i\gamma_0\gamma_3 p_3.$$

Translating this into our new Dirac formalism, we obtain:

$$\hat{\sigma}_1 = -\mathbf{i}; \quad \hat{\sigma}_2 = -\mathbf{j}; \quad \hat{\sigma}_3 = -\mathbf{k}$$

or

$$\hat{\sigma} = -\mathbf{1},$$

and

$$\boldsymbol{\gamma} = i\mathbf{1},$$

where $\mathbf{1}$ is the unit (spin) vector.

Since $\gamma_0 m = \mathbf{ik}m$ has no vector term and $\hat{\sigma}$ no quaternion, they commute, and we may derive the conventional

$$[\hat{\sigma}, \gamma_0 m] = 0$$

and

$$[\hat{\sigma}, \mathcal{H}] = [\hat{\sigma}, i\gamma_0\boldsymbol{\gamma}\cdot\mathbf{p}].$$

Now,

$$i\gamma_0\boldsymbol{\gamma}\cdot\mathbf{p} = -\mathbf{j}(\mathbf{i}p_1 + \mathbf{j}p_2 + \mathbf{k}p_3).$$

So,

$$[\hat{\sigma}, \mathcal{H}] = 2\mathbf{j}(\mathbf{ij}p_2 + \mathbf{ik}p_3 + \mathbf{ji}p_1 + \mathbf{jk}p_3 + \mathbf{ki}p_1 + \mathbf{kj}p_2)$$
$$= 2\mathbf{ij}(\mathbf{k}(p_2 - p_1) + \mathbf{j}(p_1 - p_3) + \mathbf{i}(p_3 - p_2))$$
$$= 2\mathbf{ij}\,\mathbf{1}\times\mathbf{p}.$$

In more conventional terms,

$$[\hat{\sigma}, \mathcal{H}] = 2\,\mathbf{iki}(\mathbf{k}(p_2 - p_1) + \mathbf{j}(p_1 - p_3) + \mathbf{i}(p_3 - p_2))$$
$$= 2\,\mathbf{ik}\,\boldsymbol{\gamma}\times\mathbf{p}$$
$$= 2\,\gamma_0\,\boldsymbol{\gamma}\times\mathbf{p}.$$

The factor 2 appears as a result of noncommutation. Specifically, it is the anticommuting property of the multivariate vectors in the $\gamma$ matrices which produces the doubling effect. This is the result we wished to achieve. The rest of the derivation is purely formal, and can be done either conventionally or in the new formalism. If **L** is the orbital angular momentum $\mathbf{r}\times\mathbf{p}$,

$$[\mathbf{L}, \mathcal{H}] = [\mathbf{r}\times\mathbf{p}, i\gamma_0\boldsymbol{\gamma}\cdot\mathbf{p} + \gamma_0 m]$$
$$= [\mathbf{r}\times\mathbf{p}, i\gamma_0\boldsymbol{\gamma}\cdot\mathbf{p}].$$

Taking out common factors,

$$[\mathbf{L}, \mathcal{H}] = i\gamma_0[\mathbf{r}, \boldsymbol{\gamma}\cdot\mathbf{p}]\times\mathbf{p}$$
$$= -\mathbf{ki}[\mathbf{r}, \mathbf{1}\cdot\mathbf{p}]\times\mathbf{p}$$
$$= -\mathbf{j}[\mathbf{r}, \mathbf{1}\cdot\mathbf{p}]\times\mathbf{p}.$$

Now,

$$[\mathbf{r}, \mathbf{1}\cdot\mathbf{p}]\psi = -i\mathbf{i}\left(x\frac{\partial\psi}{\partial x} - \frac{\partial(x\psi)}{\partial x}\right) - i\mathbf{j}\left(y\frac{\partial\psi}{\partial y} - \frac{\partial(y\psi)}{\partial y}\right) - i\mathbf{k}\left(z\frac{\partial\psi}{\partial z} - \frac{\partial(z\psi)}{\partial z}\right)$$
$$= i\mathbf{1}\,\psi.$$

Hence,

$$[\mathbf{L}, \mathcal{H}] = -\mathbf{ij}\,\mathbf{1}\times\mathbf{p}.$$

This, again, can be converted into conventional terms:

$$[\mathbf{L}, \mathcal{H}] = i k i \mathbf{1} \times \mathbf{p}$$
$$= i \gamma_0 \boldsymbol{\gamma} \times \mathbf{p}.$$

Using these equations, we derive

$$[\mathbf{L} - \mathbf{1}/2, \mathcal{H}] = 0$$
or
$$[\mathbf{L} + \hat{\boldsymbol{\sigma}}/2, \mathcal{H}] = 0.$$

Hence, $(\mathbf{L} - \mathbf{1}/2)$ or $(\mathbf{L} + \hat{\boldsymbol{\sigma}}/2)$ is a constant of the motion. The important aspect of this derivation is that the factor 2 is introduced as a result of anticommutation in the products of multivariate momentum operators.

## 10 The Schrödinger equation

Now, it might be assumed that the spin term ($\hat{\boldsymbol{\sigma}}/2$) is introduced with the relativistic aspect of the Dirac equation. However, using the multivariate vectors, we can obtain effectively the same result using the non-relativistic Schrödinger equation,

$$(E - V)\psi = \frac{p^2}{2m}\psi,$$

by deriving the anomalous magnetic moment of the electron in the presence of a magnetic field $\mathbf{B}$.[37] Spin, in fact, is purely a property of the multivariate nature of the $\mathbf{p}$ term, and has nothing to do with whether the equation used is relativistic or not. It is significant here that the standard derivation of the Schrödinger equation begins with the classical expression for kinetic energy, $p^2/2m = mv^2/2$.

$$T = (E - V) = \frac{p^2}{2m},$$

followed by substitution of the quantum operators $E = i \partial/\partial t$ and $p = -i\nabla$, acting on the wavefunction $\psi$, for the corresponding classical terms, to give:

$$(E - V)\psi = -\frac{1}{2m}\nabla^2\psi$$
or
$$i\frac{\partial \psi}{\partial t} - V\psi = -\frac{1}{2m}\nabla^2\psi,$$

in the time-varying case.

Now, it is possible to show that the Schrödinger equation is effectively a limiting approximation to the bispinor form of the Dirac equation in the relativistic limit. In principle, this should mean that the spin ½ term that arises from the Dirac equation has nothing to do with the fact that the equation is relativistic, but is a result of the fundamentally multivariate nature of its use of the momentum operator, equivalent to the use of Pauli matrices. In principle, we should be able to show that no new information concerning the factor 2 is introduced with special relativity. We take the Dirac equation in the form

$$(i\boldsymbol{\gamma}\cdot\mathbf{p} + m - \gamma_0 E)\psi = 0$$

and choose, without loss of generality, the momentum direction $\mathbf{i}p_x = \mathbf{p}$. Here again, also, $E$ and $\mathbf{p}$ represent the quantum differential operators, rather than their eigenvalues. This time, we make the conventional choices of matrices for $\beta$:

$$\begin{pmatrix} 1 & 0 & 0 & 0 \\ 0 & 1 & 0 & 0 \\ 0 & 0 & -1 & 0 \\ 0 & 0 & 0 & -1 \end{pmatrix}$$

and for $\gamma^1$:

$$\begin{pmatrix} 0 & 0 & 0 & -i \\ 0 & 0 & -i & 0 \\ 0 & i & 0 & 0 \\ i & 0 & 0 & 0 \end{pmatrix}$$

leading to the representation:

$$\begin{pmatrix} E-m & 0 & 0 & -p \\ 0 & E-m & -p & 0 \\ 0 & p & -E-m & 0 \\ p & 0 & 0 & -E-m \end{pmatrix} \begin{pmatrix} \psi_1 \\ \psi_2 \\ \psi_3 \\ \psi_4 \end{pmatrix} = 0.$$

This can be reduced to the coupled equations:

$$(E-m)\varphi = p\chi,$$

and

$$(E+m)\chi = p\varphi,$$

where the bispinors are given by

$$\varphi = \begin{pmatrix} \psi_1 \\ \psi_2 \end{pmatrix},$$

and

$$\chi = \begin{pmatrix} \psi_3 \\ \psi_4 \end{pmatrix}$$

Then, assuming the non-relativistic approximation $E \approx m$, for low $\mathbf{p}$, we obtain

$$\chi \approx \frac{p}{2m} \varphi$$

and

$$(E - m)\varphi = \frac{p^2}{2m}\varphi.$$

Using the same approximation, $\varphi$, here, also becomes $\psi$. Conventionally, of course, the Schrödinger equation excludes the mass energy $m$ from the total energy term $E$, and, in the presence of a potential energy $V$, we obtain:

$$(E - V)\psi = \frac{p^2}{2m}\psi.$$

## 11 Electron spin from the Schrödinger equation

Here, it will be seen that the factor 2 in the classical potential energy expression ultimately carries over into the same factor in the spin term for the electron. In our operator notation, the Schrödinger equation, whether field-free or in the presence of a field with vector potential $\mathbf{A}$, can be written in the form,

$$2mE\psi = \mathbf{p}^2\psi$$

Using a multivariate, $\mathbf{p} = -i\nabla + e\mathbf{A}$, we derive:

$$2mE\psi = (-i\nabla + e\mathbf{A})(-i\nabla + e\mathbf{A})\psi$$

$$= (-i\nabla + e\mathbf{A})(-i\nabla\psi + e\mathbf{A}\psi)$$

$$= -\nabla^2\psi - ie(\nabla.\psi\mathbf{A} + i\nabla\psi \times \mathbf{A} + \mathbf{A}.\nabla\psi + i\mathbf{A}\times\nabla\psi) + e^2\mathbf{A}^2\psi$$

$$= -\nabla^2\psi - ie(\nabla.\psi\mathbf{A} + 2\mathbf{A}.\nabla\psi + i\psi\nabla\times\mathbf{A}) + e^2\mathbf{A}^2\psi$$

$$= -\nabla^2\psi - ie(\psi\nabla.\mathbf{A} + 2\mathbf{A}.\nabla\psi) + e^2\mathbf{A}^2\psi + e\mathbf{B}\psi$$

$$= (-i\nabla + e\mathbf{A}).(-i\nabla + e\mathbf{A})\psi + e\mathbf{B}\psi$$

$$= (-i\nabla + e\mathbf{A}).(-i\nabla + e\mathbf{A})\psi + 2m\boldsymbol{\mu}.\mathbf{B}$$

This is the conventional form of the Schrödinger equation in a magnetic field for spin up, and it is the 2$m\mu.\mathbf{B}$ term which is responsible for the electron's anomalous magnetic moment. The wavefunction can be either scalar or nilpotent. Reversing the (relative) sign of $e\mathbf{A}$ for spin down, we obtain

$$2mE\psi = (-i\nabla - e\mathbf{A})(-i\nabla - e\mathbf{A})\psi$$

$$= (-i\nabla - e\mathbf{A})(-i\nabla - e\mathbf{A})\psi - 2m\mu.\mathbf{B}.$$

We can see from this derivation that the factor 2 is both introduced with the transition in the Schrödinger equation from the classical kinetic energy term, and, at the same time, produced by the anticommuting nature of the momentum operator.

## 12 The Heisenberg uncertainty principle

It is precisely because the Schrödinger equation is derived via a kinetic energy term that this factor enters into the expression for the spin, and this process is essentially the same as the process which, through the anticommuting quantities of the Dirac equation, makes ($\mathbf{L} + \hat{\sigma}/2$) a constant of the motion. Anticommuting operators also introduce the factor 2 in the Heisenberg uncertainty relation for the same reason, and the Heisenberg term relates directly to the zero-point energy derived from the kinetic energy of the harmonic oscillator. The formal derivation of the Heisenberg uncertainty relation assumes a state represented by a state vector $\psi$ which is an eigenvector of the operator $P$. In this case, the expectation value of the variable $p^2$ becomes

$$<p^2> = \psi^* P^2 \psi$$

and the mean squared variance

$$(\Delta p)^2 = \psi^*\{P - <p>I\}^2\psi = \psi^* P'^2 \psi$$

if $P' = P - <p>I$ and $I$ is a unit matrix. Similarly, for operator $Q$,

$$(\Delta q)^2 = \psi^*\{Q - <q>I\}^2\psi = \psi^* Q'^2 \psi.$$

Since $P'\psi$ and $Q'\psi$ are vectors,

$$(\Delta p)^2(\Delta q)^2 = (\psi^* P'^2 \psi)(\psi Q'^2 \psi) \geq (\psi^* P'Q'\psi)(\psi^* Q'P'\psi)$$
$$\geq |(1/2)(\psi^* P'Q'\psi - \psi^* Q'P'\psi)|^2$$
$$\geq (1/4)|(\psi^*(P'Q' - Q'P')\psi|^2$$
$$\geq (1/4)[P,Q]^2$$

Hence

$$(\Delta p)(\Delta q) \geq (1/2)[P,Q]$$
$$\geq \hbar/2$$

if $P$ and $Q$ do not commute. The significant aspect of this proof is that the factor 2 in the expression $\hbar/2$ comes from the non commutation of the $p$ operator.

## 13 The harmonic oscillator

The factor 2 in the quantum harmonic oscillator is clearly derived from the fact that the *varying* potential energy term added to the Hamiltonian, $m\omega^2 x^2/2$, is taken from a classical term of the $mv^2/2$ type. So, the Schrödinger equation for the eigenfunction $u_n(x)$ and eigenvalue $E_n$, with the $\hbar^2$ explicitly included and the spatial dimensions reduced to the linear $x$, becomes:

$$\left(-\frac{\hbar^2}{2m}\frac{\partial^2 u_n(x)}{\partial x^2} + \frac{m\omega^2 x^2}{2}\right) u_n(x) = E_n u_n(x).$$

This equation, as solved in standard texts on quantum mechanics, produces a ground state energy $\hbar\omega/2$, with the factor 2 originating in the $2m$ in the original equation. We define the new variables

$$y = \left(\frac{m\omega}{\hbar}\right)^{1/2} x \quad \text{and} \quad \varepsilon_n = E_n/\hbar\omega,$$

and the equation now becomes:

$$\left(\frac{\partial^2}{\partial^2 y} - y^2\right) u_n(y) = \left(\frac{\partial}{\partial y} + y\right)\left(\frac{\partial}{\partial y} - y\right) u_n(y) + u_n(y) \quad (1)$$

$$= \left(\frac{\partial}{\partial y} - y\right)\left(\frac{\partial}{\partial y} + y\right) u_n(y) - u_n(y) = -2\varepsilon_n u_n(y). \quad (2)$$

From this we derive

$$\left(\frac{\partial}{\partial y} - y\right)\left(\frac{\partial}{\partial y} + y\right)\left(\frac{\partial}{\partial y} - y\right) u_n(y) = (-2\varepsilon_n - 1)\left(\frac{\partial}{\partial y} - y\right) u_n(y) \quad (3)$$

and

$$\left(\frac{\partial}{\partial y} + y\right)\left(\frac{\partial}{\partial y} - y\right)\left(\frac{\partial}{\partial y} + y\right) u_n(y) = (-2\varepsilon_n + 1)\left(\frac{\partial}{\partial y} + y\right) u_n(y). \quad (4)$$

From (3), we may derive either $(\partial/\partial y - y) u_n(y) = 0$, which produces a divergent solution, or $(\partial/\partial y - y) u_n(y) = u_{n+1}(y)$ (say), which means that

$$\left(\frac{\partial}{\partial y} - y\right)\left(\frac{\partial}{\partial y} + y\right) u_{n+1}(y) = (-2(\varepsilon_n + 1) - 1) u_{n+1}(y),$$

which is (2) for $u_{n+1}$ if

$$\varepsilon_n + 1 = \varepsilon_{n+1}.$$

From (4), we may obtain either $(\partial/\partial y + y) u_n(y) = 0$, which gives us the ground state eigenfunction, $u_0(y) = \exp(-y^2/2)$; or $(\partial/\partial y - y) u_n(y) = u_{n-1}(y)$ (say). In the latter case, (4) becomes

$$\left(\frac{\partial}{\partial y} + y\right)\left(\frac{\partial}{\partial y} - y\right) u_{n-1}(y) = (-2(\varepsilon_n - 1) - 1) u_{n-1}(y)$$

if

$$\varepsilon_{n-1} = \varepsilon_n - 1,$$

which gives us a discrete series of energies $E_n$ at $n\hbar\omega$ above the ground state. From the ground state eigenfunction and (1), we obtain

$$2\varepsilon_0 - 1 = 0,$$

which gives us the ground state or 'zero-point' energy

$$E_0 = \frac{\hbar\omega}{2}.$$

Here, we can derive the factor 2 in $E_0$ directly from the introduction into Schrödinger equation of the classical term $m\omega^2 x^2/2$, which is equivalent to $mv^2/2$.

## 14 The Klein-Gordon equation

From both Dirac and Schrödinger equations, we see that fermions have half-integral spins. How, then, do we explain the integral spins of bosons, such as the photon? The answer here is that, while the fermion equation is the kinetic energy equation of Schrödinger or Dirac, based on $mv^2/2$ or $p^2/2m$, the boson equation is the potential energy equation, based on $E = mc^2$, where $m$ is now the 'relativistic', rather than the rest mass. Once again, the origin of the factor 2 is seen in the virial relation between kinetic and potential energies. The Klein-Gordon equation, which applies in quantum mechanics to the photon, derives its integral spin values from the fact that its energy term contains unit values of the mass $m$. To derive this equation, we quantize the classical relativistic energy-momentum equation,

$$E^2 - p^2c^2 - m^2c^4 = 0,$$

directly, to obtain

$$\frac{\partial^2 \psi}{\partial t^2} - \nabla^2 \psi = m^2 \psi,$$

in units where $\hbar = c = 1$. In the nilpotent algebra, the Klein-Gordon equation automatically applies to fermions, as well as to bosons, because it simply involves pre-multiplication of zero by a nilpotent differential operator. Essentially, we take the Dirac equation

$$\left(i k \frac{\partial}{\partial t} \pm i \nabla \pm i j m_0 \right)\psi = 0,$$

where

$$\psi = (kE \pm i i \mathbf{p} \pm i j m_0) e^{-i(Et - \mathbf{p} \cdot \mathbf{r})},$$

and pre-multiply by $\left(i k \frac{\partial}{\partial t} \pm i \nabla \pm i j m_0 \right)$ to give

$$\left(i k \frac{\partial}{\partial t} \pm i \nabla \pm i j m_0 \right)\left(i k \frac{\partial}{\partial t} \pm i \nabla \pm i j m_0 \right)\psi = 0,$$

or

$$\left(\frac{\partial^2}{\partial t^2} - \nabla^2 - m_0\right)\psi = 0.$$

## 15 Relativistic mass and rest mass

In principle, the kinetic energy relation is used when we consider a particle as an object in itself, described by a rest mass $m_0$, undergoing a continuous change. The potential energy relation is used when we consider a particle with its 'environment', with 'relativistic mass', in an equilibrium state requiring a discrete transition for any change. The existence of these two conservation of energy approaches has very profound implications, and arises from a very deep stratum in physics. Kinetic energy may be associated with rest mass, because it cannot be defined without it – one could consider light 'slowing down' in a gravitational field as effectively equivalent to adopting a rest mass, and, of course, photons do acquire 'effective masses' in condensed matter. Potential energy is associated with 'relativistic' mass because the latter is *defined* through a potential energy-type term ($E = mc^2$), light in free space being the extreme case, with no kinetic energy/rest mass, and 100 percent potential energy/relativistic mass. The description, in addition, seems to fit in with the halving that goes on, for a material particle, when we expand its relativistic mass-energy term ($mc^2$) to find its kinetic energy ($mv^2 / 2$). One way of looking at it is to take the relativistic energy conservation equation

$$E^2 - p^2 c^2 - m_0^2 c^4 = 0.$$

We can take regard this as a 'relativistic' mass (potential energy) equation of the form $E = mc^2$ (treating at one go the particle interacting with its environment), and proceed to quantize to a Klein-Gordon equation, with integral spin. Alternatively, we can separate out the kinetic energy term using the rest mass $m_0$. From

$$E^2 = m_0^2 c^4 \left(1 - \frac{v^2}{c^2}\right)^{-1},$$

we take the square root, and obtain

$$E = m_0 c^2 + \frac{m_0 v^2}{2} + \ldots.$$

The Schrödinger equation, of course, arises from this approach, quantizing $mv^2/2$ in the form $p^2/2m$. Now, as we have shown, using a multivariate form of the momentum operator, $\mathbf{p} = -i\nabla + e\mathbf{A}$, the Schrödinger equation produces the magnetic moment of the electron, with the required half-integral value of spin, the ½ coming from the term $mv^2/2$ or $p^2/2m$; and it is also effectively a limiting approximation to the bispinor form of the Dirac equation. In principle, as we have seen, this means that the spin ½ term that arises from the Dirac equation has nothing to do with the fact that the equation is relativistic, but arises from the fundamentally multivariate nature of its use of the momentum operator. We can now see that it comes from the very act of square rooting the energy equation in the same way as that operation produces $mv^2/2$ in the relativistic expansion. The ½ is, in essence, a statement of the act of square-rooting, which is exactly what happens when we split 0 into two nilpotents; the ½ in the Schrödinger approximation is a manifestation of this which we can trace through the ½ in the relativistic binomial approximation.

Significantly, the origin of the factor 2 is seen here in the process which square roots the expression $E^2 - m^2$. The origin of the same factor in the derivation of spin from the Dirac equation, is seen in the behaviour of the anticommuting terms which result from this process. In fact, the two factors have precisely the same origin.

Another aspect of the process is that *dimensionality*, in general, introduces two orders of meaning in a parameter – of the value (as in length/time or charge/mass), and of the squared value (as in Pythagorean/vector addition of space dimensions, or space and time, or energy and momentum, or charges / masses 'interacting' to produce forces). In a sense we are doing this with fermion and boson wavefunctions, one type being a 'square root' of the other.

## 16 Zero point energy

The importance of the factor 2 in all our examples lies in the fact that it relates together two parallel but almost independent streams of physics: the continuous and the discontinuous. Expressions involving half units of $\hbar$ do not suggest that there is such a thing as half a photon, but represent, rather, an average or integrated increase from 0 to $\hbar$. The half-values are characteristic of the continuous option in physics, the integral ones of the discontinuous option. Schrödinger, for example, represents the former, with gradualistic energy exchange and a kinetic energy equation, Heisenberg the latter, with abrupt transitions between states in integer values of $\hbar\omega$, determined by bosonic (potential energy) equations. Both approaches are equally valid, although they represent divergent physical models, and it is not surprising that a completely continuous theory of stochastic electrodynamics, based on the existence of zero-point energy of value $\hbar\omega/2$, at each point in space, has developed as a rival to the purely discrete theory of the quantum with energy $\hbar\omega$.

Stochastic electrodynamics has been successful in providing classical explanations of the Planck black body radiation law from equipartition, and of Bose-Einstein statistics for photons.[38-42] In addition, spontaneous emission, Bohr transitions, *zitterbewegung*, Van der Waals forces, and the third law of thermodynamics, have been shown as classical phenomena arising from electromagnetic radiation, with the stochastic energy spectrum of $\hbar\omega/2$ per normal mode of vibration,[38,43] and the same principle has been used to derive the Schrödinger equation from Newtonian mechanics.[44,45] Stochastic electrodynamics appears to form a successful continuous option to discrete quantum mechanics based on the use of a half value of the energy quantum.

In fact, not only are both discrete and continuous options possible – both are required within a system. Discrete systems have to incorporate continuity, and continuous ones discreteness. Schrödinger thus has a continuous system based on $\hbar/2$, but incorporates discreteness (based on $\hbar$) in the process of measurement – the so-called collapse of the wavefunction. Heisenberg, on the other hand, has a discrete system, based on $\hbar$, but incorporates continuity (and $\hbar/2$) in the process of measurement – via the uncertainty principle and zero-point energy. Continuity and discontinuity must both be present in a successful system, so whichever is not present in the mathematical structure must be introduced in the process of measurement. In addition, it would seem, nature always manages to provide a route by which $\hbar\omega/2$ in one context becomes $\hbar\omega$ in another. This occurs, for example, in the case of black-body radiation, where the spontaneous emission of energy of value $\hbar\omega$ is produced by the combined effect of the $\hbar\omega/2$ units of energy provided by both oscillators and zero-point field.[46]

Just as the relativistic expression for kinetic energy presents a problem in the asymptotic approach to classical conditions, so does Planck's quantum law of black body radiation. As Einstein and Stern noticed in 1913, the Planck equation for the energy of each oscillator

$$U = \frac{h\nu}{\exp(h\nu/kT) - 1}$$

does not reduce to the classical limit $kT$ when $kT \gg h\nu$, but to $kT - h\nu/2$. Planck himself, in his so-called 'second theory of radiation', based on discrete emission but classical or continuous absorption, had obtained the modified law

$$U = \frac{h\nu}{\exp(h\nu/kT) - 1} + h\nu/2,$$

which suggested, as he said, that, even at the absolute zero of temperature, each oscillator had an energy equivalent to $h\nu/2$ at each frequency $\nu$.

In quantum mechanics, as we have seen, the zero-point energy term, $h\nu/2$ or $\hbar\omega/2$, is derived from the harmonic oscillator solution of the Schrödinger equation. In the Heisenberg formulation it appears as a result of the $\hbar/2$ term involved in the uncertainty principle. The derivation via Schrödinger shows the kinetic origins of the factor 2. The derivation from the uncertainty principle suggests the origin of this fundamental constant in continuum physics, as opposed to the constant $h$ used in discrete theories. It certainly does not suggest that there is any such thing as a half-photon!

## 17 Radiation reaction

The $\hbar\omega/2 \rightarrow \hbar\omega$ transition for black body radiation can also be seen in terms of radiation reaction. Perhaps, surprisingly, this has an intimate connection with the distinction between the relativistic and rest masses of an object. The act of defining a rest mass also defines an isolated object, and one cannot define *kinetic* energy in terms of anything but this rest mass. If, however, we take a *relativistic* mass, we are already incorporating the effects of the environment. The most obvious instance is that of the photon. The photon has no rest mass, only a relativistic mass; $mc^2$ for a photon behaves exactly like a classical potential energy term, as well as having the exact form of a potential energy for a body of mass $m$ and speed $c$. A particular instance we have used is the application of a material gas analogy to a photon gas in producing radiation pressure $\rho c^2/3$. Action and reaction occurs in this instance because the doubling of the value of the energy term comes from the doubling of the momentum produced by the rebound of the molecules/photons from the walls of the container – a classic two-step process, like the two-way speed of light.

The energy involved in both material and photon gas pressure derivations is clearly a potential energy term (the material gas energy having to be halved to relate the kinetic energy of the molecules to temperature), and its double nature is derived from the two-way process which it involves, which is the same thing as saying that it is Newton's action *and* reaction. The same thing happens with radiation reaction, which produces a 'mysterious' doubling of energy $h\nu/2$ to $h\nu$ in many cases (and also *zitterbewegung* for the electron, which is interpreted as a switching between two states). In another context, Feynman and Wheeler also produce a doubling of the contribution of the retarded wave in electromagnetic theory, at the expense of the advanced wave, by assuming that the vacuum behaves as a perfect absorber and reradiator of radiation. In principle, this seems to be equivalent to assuming a filled vacuum for advanced waves (equivalent to Dirac's filled vacuum for antimatter), and relates to previously stated ideas that continuity of mass-energy in the vacuum is related to the unidirectionality of time.[47-50] There are also connections with some paradoxes in special relativity.

## 18 Paradoxes in special relativity

As we have seen, we frequently find the factor 2 where we need to introduce such ideas as radiation reaction in the theory of zero point energy. Incorporating radiation reaction means that we are also incorporating the effect of Newton's third law, the process which produces the required doubling in the case of material and photon gases, and other steady-state processes. However, many of the same results, as in the anomalous magnetic moment of the electron, are also explained by special relativity. It has been argued by C. K. Whitney that the correct result for the electron is obtained by treating the transmission of light as a two-step process involving absorption and emission.[51] This is interesting because it is equivalent to incorporating both action and reaction, or the potential energy equation, and the same result follows classically by defining the potential energy at the moment the field is switched on. However, if we use kinetic energy, or a one-step process, we also need relativity, because, once we introduce rest mass, we can no longer use classical equations. ('Relativistic mass' is, of course, specifically *designed* to preserve classical energy conservation!) The two-step process is analagous to the use of radiation reaction, so it follows, in principle, that a radiation reaction is equivalent to adding a relativistic 'correction' (such as the Thomas precession).

Whitney's argument that the two-step process removes those special relativistic paradoxes which involve apparent reciprocity, is also interesting, because special relativity, by including only one side of the calculation, effectively removes reciprocity, and so leads to such things as asymmetric ageing in the twin paradox. The argument, put forward by some authors,[52] that the problems arise in Einstein's denial of the aether may also be relevant if we translate it to the vacuum, because no vacuum means no 'environment', and, therefore, no 'reaction'. Similar arguments again apply to the idea that the problem lies in attempting to define a one-way speed of light that cannot be measured, because a two-way speed measurement of the speed of light also requires a two-step process.

Whitney further shows that the classic light-bending and perihelion precession 'tests' of General Relativity can be derived using a two-step process. This, again, is of interest, because, as shown here, it is certainly possible to derive the light bending by classical arguments using kinetic energy (which is the same thing as using special relativity, because light has no rest mass), and it is also possible to derive perihelion precession using special relativity, as a number of authors have demonstrated.[14,16,18]

## 19 The Jahn-Teller effect

From the earlier sections, it would appear that all the important factors of 2 in classical physics, relativity and quantum physics, result from a choice between using kinetic or potential energies, and that this is equivalent to using either the action side, or the combined action and reaction sides, of Newton's third law of motion. This, in turn, derives from a choice between using continuous or discrete solutions, or changing or fixed ones. A series of further arguments show that the origin of the factor lies in the symmetry between the action of an object and the reaction of its environment – which may be either material or vacuum.[33] A fermionic object on its own shows changing behaviour, requiring an integration which generates a factor ½ in the kinetic energy term, and a sign change when it rotates through $2\pi$, but a conservative 'system' of object plus environment shows unchanging behaviour, requiring a potential energy term, which is twice the kinetic energy.

This kind of argument makes sense of the boson / fermion distinction and the spin 1/½ division between the particle types in a fundamental way, as well as leading to supersymmetry, vacuum polarization, pair production, renormalization, and so on, because the halving of energy in 'isolating' the fermion from its vacuum or material 'environment' is the same process as mathematically square-rooting the quantum operator via the Dirac equation. Bell *et al* have shown that integral spins are automatically produced from half-integral spin electrons using the Berry phase, and, by generalizing this kind of result to all possible environments, we may extend the principle in the direction of supersymmetry.[53] In principle, we propose that energy principles determine that *all* fermions, in whatever circumstances, may be regarded either as isolated spin ½ objects or as spin 1 objects in conjunction with some particular material or vacuum environment, or, indeed, the 'rest of the universe'.

While hypothetically isolated fermions may follow the Dirac equation, derived from the kinetic energy relation, and similarly isolated bosons follow the Klein-Gordon equation, derived from the potential energy relation, the same particles in real situations behave very differently. Fermions with spin ½ become spin 1 particles when taken in conjunction with their environment, whatever that may be. The Jahn-Teller effect and Aharanov-Bohm effect are examples. Treated semi-classically, the Jahn-Teller effect, for electrons in condensed matter, couples the factors associated with the motions of the relevant electronic and nuclear coordinates so that different parts of the total wavefunction change sign in a coordinated manner to preserve the single-valuedness of the total wavefunction. This is possible because the time-scale of the nuclear motions is much greater than that for the electronic transitions. Neither the nuclear nor the electronic wavefunction are single-valued by themselves, but the total wavefunction becomes so through the Jahn-Teller effect.

In more general terms, the relationship between a fermion and 'the rest of the universe' can be considered as similar to the that of the total wavefunction in the Jahn-Teller effect. Isolated fermions cannot have single-valued wavefunctions, but the *total* wavefunction representing fermion plus 'rest of the universe' must be single-valued. This duality occurs with the actual creation of the fermion state. To split away a fermion from a 'system' (or 'the universe'), we have to introduce a coupling as a mathematical description of the splitting. The coupling to the rest of the universe preserves the single-valued nature of the total wavefunction, automatically introducing the extra term known as the Berry phase. Many physical effects, including the Aharanov-Bohm effect, as well as the Jahn-Teller effect, are already associated with this phase, and there are, no doubt, many others waiting to be discovered.

The reverse effect must also exist, in which bosons of spin 0 or 1 couple to an 'environment' to produce fermion-like states. Perhaps the Higgs mechanism occurs in this way, but a more immediate possibility is the coupling of gluons to the quark-gluon plasma to deliver the total spin of ½ or 3/2 to a baryon. The six-component baryon wavefunction has states equivalent to ($kE \pm ii p_x + ijm$)($kE \pm ii p_y + ijm$)($kE \pm ii p_z + ijm$), where the $p_x$, $p_y$, $p_z$ and $\pm$ represent the six degrees of freedom for **p**.[33] These, of course, exist simultaneously in a gauge-invariant state, but we can imagine the **p** rotating through the three spatial positions leaving terms like ($kE \pm ii p + ijm$)($kE + ijm$)($kE + ijm$); ($kE + ijm$)($kE \pm ii p + ijm$)($kE + ijm$), with the gluons 'transferring' the **p** between one ($kE + ijm$) and another, and so becoming bosons of spin 1 with an effective contribution from the 'environment' due to the gluon sea making them transfer spin ½.

It is almost certainly a universal principle that fermions/bosons always produce a 'reaction' within their environment, which couples them to the appropriate wavefunction-changing term, so that the potential / kinetic energy relation can be maintained at the same time as its opposite. We can relate this to the whole process of renormalization which produces an infinite chain of such couplings through the vacuum. The coupling of the vacuum to fermions generates 'boson-images' and vice versa. This suggests that the loop diagrams that lead to renormalisation could produce the required cancellation of fermion with boson loops without requiring the existence of extra boson or fermion equivalents.[54]

## 20 Renormalization

To understand the principle, we need to use the nilpotent version of the Dirac wavefunction, which is, typically, ($kE + ii\mathbf{p} + ijm$) for a fermion and ($-kE + ii\mathbf{p} + ijm$) for an antifermion, these being abbreviated representations of 4-term bra and ket vectors, cycling through the full range of $\pm E$ and $\pm\mathbf{p}$ values. In terms of the 'environment' principle, a fermion generates an infinite series of interacting terms of the form:

$(k E + i i \mathbf{p} + i j m)$
$(k E + i i \mathbf{p} + i j m)(- k E + i i \mathbf{p} + i j m)$
$(k E + i k \mathbf{p} + i j m)(- k E + i i \mathbf{p} + i j m)(k E + i i \mathbf{p} + i j m)$
$(k E + i i \mathbf{p} + i j m)(- k E + i i \mathbf{p} + i j m)(k E + i i \mathbf{p} + i j m)(- k E + i i \mathbf{p} + i j m)$, etc.

Selection of the appropriate terms in QED calculations now leads to a cancellation of the boson and fermion loops of opposite sign at any level. The $(k E + i i \mathbf{p} + i j m)$ and $(- k E + i i \mathbf{p} + i j m)$ vectors are an expression of the behaviour of the vacuum state, which acts like a 'mirror image' to the fermion. An expression such as

$$(k E + i i \mathbf{p} + i j m) \, k ( k E + i i \mathbf{p} + i j m)$$

is part of an infinite regression of images of the form

$$(k E + i i \mathbf{p} + i j m) \, k ( k E + i i \mathbf{p} + i j m) \, k ( k E + i i \mathbf{p} + i j m) \, k ( k E + i i \mathbf{p} + i j m)...$$

where the vacuum state depends on the operator that acts upon it, the vacuum state of $(k E + i i \mathbf{p} + i j m)$, for example, becoming $k ( k E + i i \mathbf{p} + i j m)$. In addition,

$$(k E + i i \mathbf{p} + i j m) \, k ( k E + i i \mathbf{p} + i j m) \, k ( k E + i i \mathbf{p} + i j m) \, k ( k E + i i \mathbf{p} + i j m)...$$

is the same as

$$(k E + i i \mathbf{p} + i j m)(- k E + i i \mathbf{p} + i j m)(k E + i i \mathbf{p} + i j m)(- k E + i i \mathbf{p} + i j m)....$$

So, the infinite series of creation acts by a fermion on vacuum turns out to be the mechanism for creating an infinite series of alternating boson and fermion states as required for supersymmetry and renormalization. This is only true if the series is infinite, because each 'antifermion' bracket has to be postmultiplied by $k$ to alter the sign of its $E$ term. It also requires spin terms $\mathbf{p}$ of the same sign to produce spin 1 bosons; spin 0, such as the mass-generating Higgs boson, would break the sequence.

The 'mirror imaging' process implies an infinite range of virtual $E$ values in vacuum adding up to a single finite value, exactly as in renormalisation. Significantly, the vacuum wavefunctions for the fermion and antifermion are of the complementary forms, $(- k E + i i \mathbf{p} + i j m)$ and $( k E + i i \mathbf{p} + i j m)$, to those for the particles. It is also significant that, in the classical context, the related Feynman-Wheeler process of vacuum absorption of radiation (discussed in section 17) again reduces the infinite electron self-energy to a finite mass.

## 21 Supersymmetry

'Supersymmetry' may be part of a much more general pattern. Bosons and fermions seem to require 'partner states' as much as potential and kinetic energy are needed to fully describe conservation. As previously stated, the kinetic energy relation is used when we consider a particle as an object in itself, described by a rest mass $m_0$, undergoing a continuous change. The potential energy relation is used when we consider a particle within its 'environment', with 'relativistic mass', in an equilibrium state requiring a discrete transition for any change. This fundamental relation, leads to the significant fact that the nilpotent wavefunctions, in principle, produce a kind of supersymmetry, with the supersymmetric partners not being so much realisable particles, as the couplings of the fermions and bosons to vacuum states.

The nilpotent operators defined for fermion wavefunctions are also supersymmetry operators, which produce the supersymmetric partner in the particle itself. The $Q$ generator for supersymmetry is simply the term ($kE + ii\mathbf{p} + ijm$), and its Hermitian conjugate $Q†$ is ($-kE + ii\mathbf{p} + ijm$). Written out in full, of course, these are respectively four-term bra and ket vectors, with the $E$ and $\mathbf{p}$ values going through the complete cycle of + and – values; and, with the application of the same normalization that we have used for the vacuum operator, the anticommutator of $Q$ and $Q†$ becomes effectively $E$, or the Hamiltonian, as in conventional supersymmetry theory. Multiplying by ($kE + ii\mathbf{p} + ijm$) converts bosons to fermions, or antifermions to bosons (the $\mathbf{p}$ can, of course, be + or –). Multiplying by ($-kE + ii\mathbf{p} + ijm$) produces the reverse conversion of bosons to antifermions, or fermions to bosons. In conventional supersymmetry theory, boson contributions and fermion contributions are of opposite sign (with the operators having opposite signs of $E$) and automatically cancel in loop calculations. The present theory retains this advantage without requiring *extra* (undiscovered) supersymmetric partners to the known fermions and bosons.

The spin ½ state, as we have seen, is always due to kinetic energy, implying continuous variation. and it is essentially that of the isolated fermion. Unit spin comes from the potential energy of a stable state, and represents either a boson with two nilpotents (which are not nilpotent to each other), or a bosonic-type state produced by a fermion interacting with its material environment or vacuum, and, as a consequence, manifesting Berry phase, Thomas precession, relativistic correction, radiation reaction, *zitterbewegung*, or whatever else is needed to produce the 'conjugate' environmental spin state. In the case of the isolated fermion we are treating the action half of Newton's third law; in the case of the fermion interacting with its environment, it is the action and reaction pair. The existence of 'supersymmetric' partners seemingly comes from the duality represented by the choice of fermion or fermion plus environment.

In this context it is significant that, while the Klein-Gordon equation automatically applies to fermions as well as to bosons, the Dirac equation applies to 'spin 1' particles created by the combination of fermion plus environment. The consequences are the Berry phase, the Aharonov-Bohm effect, the Jahn-Teller effect, the quantum Hall effect, *zitterbewegung*, and other such phenomena. For a fermion or boson acting in this way with its 'environment', the supersymmetric operators do not demand an *extra* set of bosons or fermions; the coupling of fundamental particles to the vacuum becomes automatic in an infinite series of entangled states.

## 22 Aharonov-Bohm effect

The Aharonov-Bohm effect can be considered as an analogue of the Jahn-Teller effect, and as another example of the effect of the Berry phase, but a consideration of this phenomenon suggests that it may lead to a more profound understanding of the meaning of the factor 2 in fundamental physics. In the Aharonov-Bohm effect, electron interference fringes, produced by a Young's slit arrangement, are shifted by half a wavelength in the presence of a solenoid whose magnetic field, being internal, does not interact with the electron but whose vector potential does. The half-wavelength shift turns out to be a feature of the topology of the space surrounding the discrete flux-lines of the solenoid. This space is not *simply-connected*, that is, a circuit round the flux line cannot be deformed continuously down to a point. Effectively, the half-wavelength shift, or equivalent acquisition by the electron of a half-wavelength Berry phase, implies that an electron path between source and slit, round the solenoid, involves a *double-circuit* of the flux line (to achieve the same phase), and a path that goes round a circuit twice cannot be continuously deformed into a path which goes round once (as would be the case in a space without flux-lines).

The presence of the flux line is equivalent, as in the quantum Hall effect and fractional quantum Hall effect, to the extra fermionic ½-spin which is provided by the electron acting in step with the nucleus in the Jahn-Teller effect and makes the potential function single-valued, and the circuit for the complete system a single loop. It is particularly significant that the $U(1)$ (electromagnetic) group responsible for the fact that the vacuum space is not simply connected is isomorphic to the integers under addition. In effect, the spin-½, ½-wavelength-inducing nature of the fermionic state (in the case of either the electron or the flux line) is a product of discreteness in both the fermion (and its charge) and the space in which it acts. (The $U(1)$ group is also relevant to fermionic states with zero electric charge, through the $SU(2) \times U(1)$ mixing; the $U(1)$ component may even be considered, in such cases, as a necessary consequence of fermionic discreteness.) In principle, the very act of creating a discrete particle requires a splitting of the continuum vacuum into *two* discrete halves (as with the bisecting of the rectangular figure with which we started), or (relating the concept of discreteness to that of dimensionality) two square roots of 0. (Mathematically, the identification of 1 as separate from 0 also implies that 1+1=2, reflecting the fact that physics and mathematics have a common origin in the process

of counting.)

## 23 Conclusion

The numerical factor 2 has become an almost universal component of fundamental physics, playing a significant role in both quantum theory and relativity. Its origin and meaning can be explained in surprisingly simple terms, using relatively unsophisticated mathematics. In fact, the origin of the factor 2, in all significant cases – classical, quantum, relativistic – is in the virial relation between kinetic and potential energies. Careful study of the factor reveals that it is the link between the continuous and discrete physical domains, and their manifestations in many areas of physics. In principle, the differences between stochastic and quantum electrodynamics, Lorentz- and Einstein-type relativities, Schrödinger and Heisenberg versions of quantum mechanics, waves and particles, spin 1/2 and spin 1 units, fermions and bosons, are nothing but those between kinetic and potential energies, between averaged-out changing and fixed steady-state values, or, indeed, between triangles and rectangles.

The result of all these cases is that kinetic energy variation may be thought of as continuous, but starting from a discrete state; potential energy variation, on the other hand, is a discrete variation, starting from a continuous state. Each creates the opposite in its variation from itself. Kinetic energy and potential energy create each other, in the same way as they are related by a numerical relationship. We can consider the kinetic energy relation to be concerned with the action side of Newton's third law, while the potential energy relation concerns both action and reaction. Because of the necessary relation between them, each of these approaches is a proper and complete expression of the conservation of energy. Ultimately, the factor 2 is an expression of the discreteness of both material particles (or charges) and the spaces between them, as opposed to the continuity of the vacuum in terms of energy. The same discreteness also implies (though more subtly) the concept of dimensionality.

In more general terms, the factor 2 is an expression of a fundamental duality in nature, and duality is the result of trying to create something from nothing – the Aharanov-Bohm effect is a classic case, as is also the nilpotent algebra used for the fermion wavefunction. Fundamentally, physics does this when it sets up a probe to investigate an intrinsically uncharacterizable nature. Nature responds with symmetrical opposites to the characterization assumed by the probe, which, in its simplest form, is constituted by a discrete point in space. It has been demonstrated previously that this generates a symmetrical group of fundamental parameters (space – the original probe – time, mass and charge – the combined response), which are defined by properties which split the parameters into three $C_2$ groupings, depending on whether they are conserved or nonconserved, real (or orderable) or imaginary (or nonorderable), continuous or discrete. Each of these divisions may be held responsible for a factor 2, for duality seems to be the necessary result of any attempt at creating singularity.

While the continuous or discrete duality is obvious from the distinction between potential and kinetic energies, this distinction also incorporates the duality between conserved and nonconserved quantities (or fixed and changing conditions). The duality may also be expressed in terms of the distinction between space-like and time-like theories (for example, those of Heisenberg and Schrödinger, or of quantum mechanics and stochastic electrodynamics), which are not only distinguished by being discrete and continuous, but also by being real and imaginary. Though a single duality separates such theories, it is open to more than one interpretation because each pair of parameters is always separated by two distinct dualities.

The very concept of duality implies that the actual process of counting is created at the same time as the concepts of discreteness, nonconservation, and orderability are separated from those of continuity, conservation, and nonorderability. The mathematical processes of addition and squaring are, in effect, 'created' at the same time as the physical quantities to which they apply. The factor 2 expresses dualities which are fundamental to the creation of both mathematics and physics.

**Appendix**

A correlation between alternative explanations for the factor 2 in various aspects of physics:

| | | | | | |
|---|---|---|---|---|---|
| Kinematics | | | | V | X |
| Gases | A | | | V | |
| Orbits | A | | | V | X |
| Radiation pressure | A | E | | V | |
| Gravitational light deflection | | | R | V | |
| Fermion/boson spin | C D | | O R | V | |
| Zero-point energy | A C | | | V | X |
| Radiation reaction | A | E | R | V | |
| SR paradoxes | A | E | | | |

| | |
|---|---|
| A | action and reaction |
| C | commutation relations |
| D | dimensionality |
| E | absorption and emission |
| O | object and environment |
| R | relativity |
| V | virial relation |
| X | continuity and discontinuity. |